\documentclass[prb,twocolumn,showpacs,preprintnumbers,superscriptaddress]{revtex4-1}
\usepackage[usenames,dvipsnames]{color}
\usepackage{amssymb} 
\usepackage{times}
\usepackage{graphicx}
\usepackage{graphicx}
\usepackage{dcolumn}
\usepackage{datetime}
\usepackage{soul}
\usepackage{subfigure}
\usepackage[bookmarks, colorlinks=true, plainpages = false,citecolor = red, urlcolor = black, 
filecolor = blue]{hyperref}
\usepackage{multirow} 
\usepackage[english]{babel}
\usepackage{graphicx}
\usepackage{amsmath}
 
\usepackage[english]{babel}
\usepackage{graphicx}
\usepackage{amsmath}
\usepackage[title,titletoc,toc]{appendix}
 
\begin{document}
\title{Valley filters, accumulators and switches induced in graphene quantum dots by lines of adsorbed hydrogen atoms}
 
\author{M. Azari} 

\author{G. Kirczenow}

\affiliation{Department of Physics, Simon Fraser
University, Burnaby, British Columbia, Canada V5A 1S6}

\date{\today}

\begin{abstract}\noindent
We present electronic structure and quantum transport calculations that predict conducting channels induced in graphene quantum dots by lines of adsorbed hydrogen atoms to function as highly efficient, experimentally realizable valley filters, accumulators and switches. The underlying physics is a novel property of graphene Dirac point resonances (DPRs) that is revealed here, namely, that an electric current passing through a DPR-mediated conducting  channel in a given direction is carried by electrons of {\em only one} of the two graphene valleys. Our predictions apply to lines of hydrogen atoms adsorbed on graphene quantum dots that are either free standing or supported on a hexagonal boron nitride substrate.
 
\end{abstract}

 
\maketitle
\section{Introduction}
Graphene is a two dimensional material whose conduction (valence) band possesses two \textit{valleys}, i.e., minima (maxima) at the same energy but with differing momenta. Consequently, graphene is attracting attention as a possible platform for valleytronics, where the electron's valley index is to be exploited as a further degree of freedom. 
In terms of the correspondence between the principal elements of spintronics and valleytronics, the potential importance of valley filters in valleytronics is evident, in view of the decisive role of spin filters in spintronics. 
Valley filtering mechanisms have been proposed based on topological considerations,\cite{XiaoPRL,XiaoReview,pan2015perfect} transport through nano-constrictions,\cite{rycerz2007valley} graphene line defects,\cite{gunlycke2011graphene,Chen14,Ingaramo,Cheng} disorder,\cite{Cresti16,An} optical\cite{abergel2009generation,Golub,Kundu,Lago2017,Qu2018} or thermal\cite{Chenthermal,Zhang2018} effects, electrostatic potentials,\cite{Costa15,wang2017valley,Ang2017} monolayer-bilayer graphene boundaries,\cite{Nakanishi} electron injection into graphene with broken inversion symmetry,\cite{george,azari2017gate} mirror symmetry breaking,\cite{Asmar} substrate defects,\cite{daCosta} p-n junctions,\cite{Sekera,Park2018} grain boundaries, \cite{Nguyen16} strains\cite{Zhangstrained,Settnesstrained,Faraj2017} nanobubbles \cite{Settnes16,Munoz2017} and SiC nanoribbon junctions \cite{Zheng2017}. Interest in this topic is intense and growing rapidly at present.\cite{XiaoPRL,rycerz2007valley,abergel2009generation,Nakanishi,XiaoReview,gunlycke2011graphene,Chen14,Ingaramo,pan2015perfect,Costa15,Chenthermal,george,azari2017gate,Nguyen16,Settnes16,Golub,Kundu,Cheng,Cresti16,An,Asmar,Munoz2017,daCosta,Zhangstrained,Settnesstrained,Sekera,wang2017valley,Lago2017,Ang2017,Faraj2017,Qu2018,Zhang2018,Park2018,Zheng2017}
\begin{figure}[b!]
\centering
\includegraphics[width=1.0\linewidth]{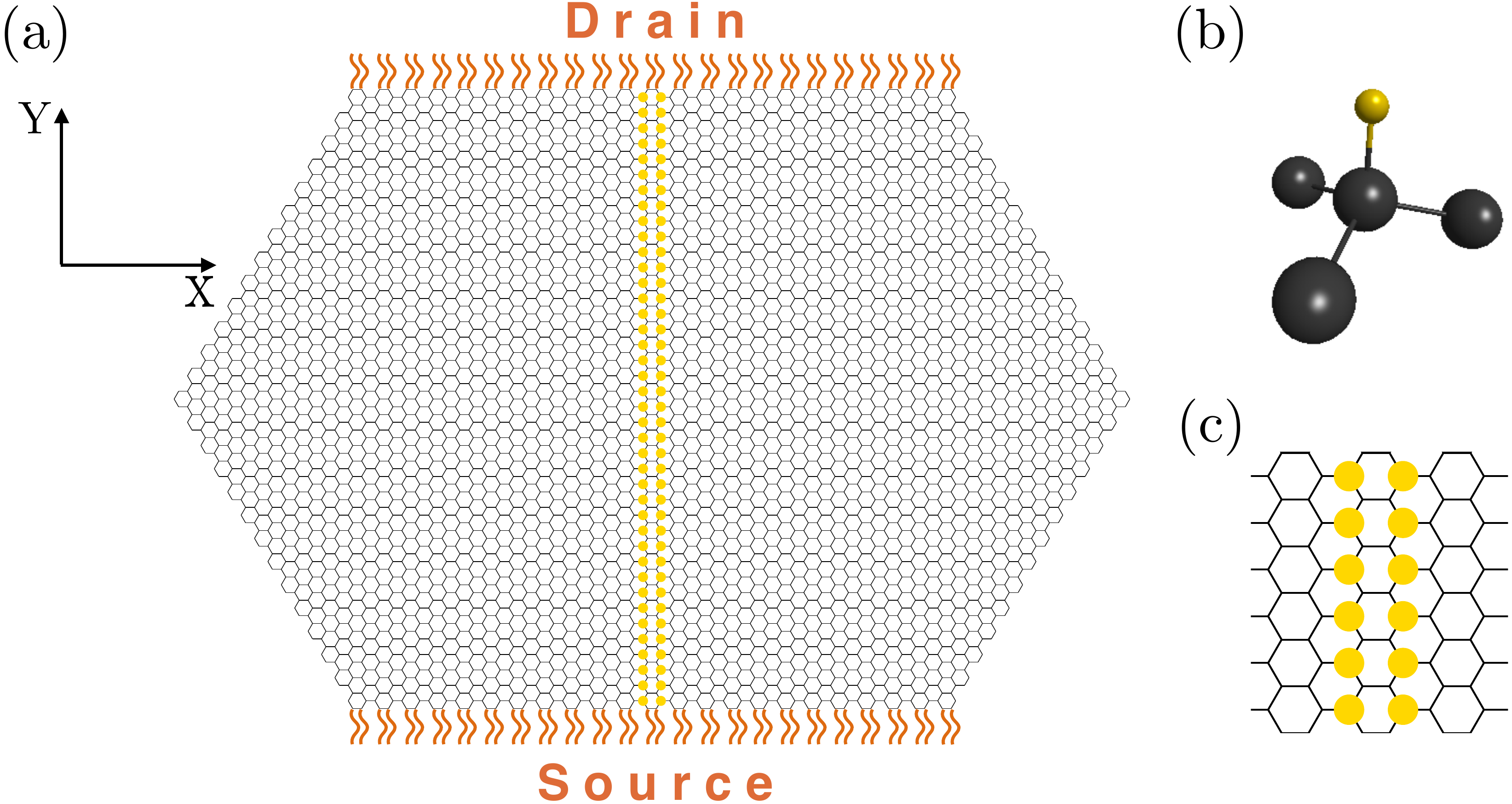}
\caption{(Color online)Fig.1. (a) 2-terminal monolayer graphene quantum dot with armchair edges. Electron source and drain contacts are each composed of 46 semi-infinite 1D leads (orange wavy lines), that connect the quantum dot to the reservoirs. Hydrogen atoms (yellow disks) are placed on top of carbon atoms and divide the nanostructure into two equal parts. (b) Relaxed geometry of adsorbed hydrogen on graphene. (c) Fragment of the double hydrogen line on graphene. The double H-line may be viewed as a series of $x$-oriented H-dimers; such para dimers on graphene are very stable\cite{Ferro2008}.      
}
\label{nanostructure} 
\end{figure} 
An obstacle to further progress is that well controlled, reproducible experimental realization of many of the proposed nanoscale devices is challenging. In this respect, nanostructures defined by arrays of atoms or molecules adsorbed on graphene may offer an attractive alternative since precisely defined patterns of adsorbed atoms can be produced to specification on surfaces with scanning tunneling microscopes.\cite{Eigler1990,Crommie1993,Hla} In particular, positioning of adsorbed hydrogen atoms on graphene with atomic precision has been demonstrated.\cite{gonzalez2016atomic}
 Adsorbed hydrogen modifies the electronic properties of graphene\cite{Sofo2007,Boukhvalov2008,Elias2009,Soriano2011} and endows it with new functionalities that may find applications in future technologies.\cite{Pumera2013,Ferrari2015,Jiang2018} However, valley filtering in graphene by means of adsorbed hydrogen (or other adsorbed species) has not been explored to date either experimentally or theoretically. Here we propose a novel, efficient, experimentally realizable mechanism of valley filtering in monolayer graphene quantum dots decorated by a double line of hydrogen atoms as depicted in Fig. \ref{nanostructure}. This new mechanism exploits scattering resonances induced in the graphene at energies near the Dirac point by adsorbed atoms\cite{Robinson08,Wehling09X,Wehling09,Wehling10,ihnatsenka2011dirac,Saffarzadeh12,ReviewRibbon,Irmer18} and other defects.\cite{Skrypnyk06,Wehling07,Pereira08,Basko08}
These ``Dirac point resonances" are due to coupling between discrete states and the graphene continuum, as in Fano interference phenomena.\cite{Fano,ihnatsenka2011dirac} For the graphene quantum dots that we consider we find the hydrogen Dirac point resonances to fall within a quantum confinement-induced energy gap. Thus the hydrogen lines behave as a quasi-1D conductor with conduction mediated by Dirac point resonances. Our quantum transport calculations reveal that electrons belonging to the two graphene valleys travel in {\em opposite} directions along this 1D conductor. Thus an electric current flowing in a given direction through this conductor is carried by electrons belonging to just one of the graphene valleys. If the direction of the electric current is reversed, conduction switches to the other valley. Furthermore the electric current results in strong valley accumulation along the hydrogen lines. The electronic states involved are localized exponentially to the vicinity of the hydrogen lines. Thus we expect impurities and other imperfections located more than a few lattice  spacings from the hydrogen lines not to degrade significantly the highly efficient operation of this novel nanoscale electrically controlled valley filter, accumulator and switch.

The paper is organized as follows. In Sec. II, we present the model that we use to describe the graphene quantum dot and adsorbed hydrogen. In Sec. III, we present the results of our calculations of the electrical conductance of the graphene quantum dot with and without adsorbed hydrogen, and with and without a symmetry-breaking substrate. In Sec. IV we present electronic band structure calculations for graphene ribbons with double lines of adsorbed hydrogen. With the help of the calculated band structures, we then interpret the results of our conductance calculations physically and relate the conductance features that we have found to valley filtering. In Sec. V we present our results for the electron valley filtering and the valley accumulations induced in the graphene dot near the hydrogen lines when an electric current flows through this system. Our conclusions are summarized in Sec. VI.
\section{The Model} 
    In Fig.\ref{nanostructure} (a), (c) the double line of hydrogen atoms (yellow disks) is oriented in the y-direction. The electron source and drain contacts that connect the quantum dot to the reservoirs are modeled as groups of semi-infinite 1D ideal leads (orange wavy lines). The geometry of a hydrogen atom adsorbed on graphene is shown in Fig.\ref{nanostructure} (b). The hydrogen and carbon atom to which hydrogen binds are 1.47 and 0.32\AA~above the graphene plane.\cite{ihnatsenka2011dirac} 
\begin{figure*}[t]
\centering
\includegraphics[width=1.0\linewidth]{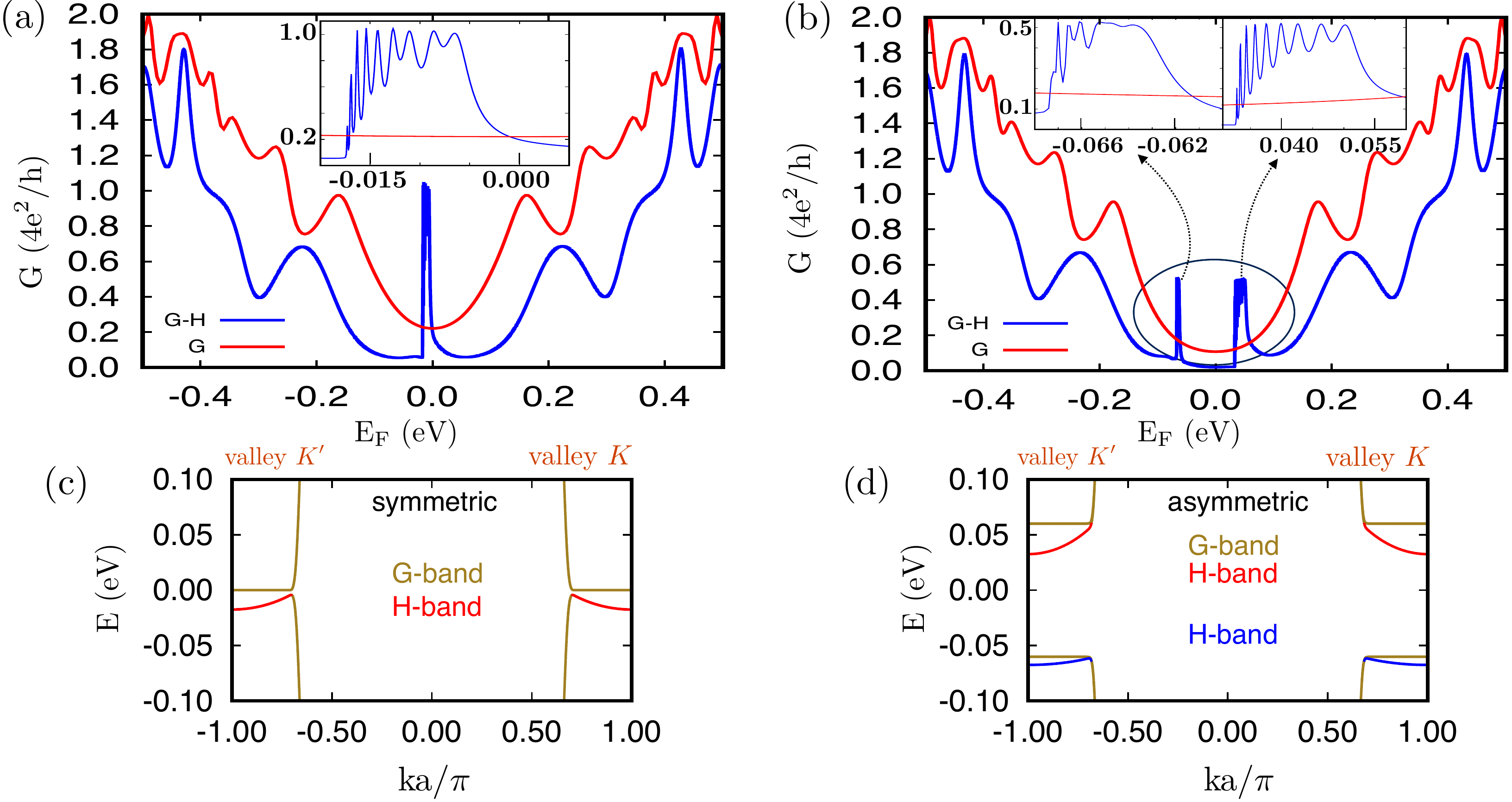}
\caption{(Color online)(a) and (b): Calculated 2-terminal conductance $G$ [Eq. \ref{conductance}] of graphene quantum dot in Fig.\ref{nanostructure} with (G-H) and without (G) hydrogen lines at zero temperature for symmetric and asymmetric carbon on-site energies vs. Fermi energy, respectively. Insets: Conductance fine structure. (c) and (d): Band structures of zigzag graphene nanoribbon with lines of hydrogen and symmetric and asymmetric on-site energies of carbon atoms, respectively. States localized near the hydrogen lines (H-bands) and other graphene states (G-bands) are present.
}
\label{conband}    
\end{figure*} 
 
 To model the nanostructure in Fig.\ref{nanostructure}(a), we have used a tight-binding Hamiltonian that accurately describes the Dirac point resonances of hydrogen on graphene:\cite{ihnatsenka2011dirac}
\begin{multline} 
H=\sum_n \Delta_n (a_n^{\dag}a_n)-\sum_{\langle n,m\rangle}t_{nm}(a_n^{\dag}a_m+h.c.)+\\
\sum_{\alpha}\epsilon_{\alpha}d_{\alpha}^{\dag}d_{\alpha}+\sum_{\alpha,n}\gamma_{\alpha,n}(d_{\alpha}^{\dag}a_n+h.c.)
\label{Hamiltonian}
\end{multline}              
The first two terms are the well known nearest-neighbour tight-binding Hamiltonian of graphene. $\Delta_n$ is the on-site energy of the carbon atoms. The third term is the on-site energy of adsorbed hydrogen atoms. The last term represents the hopping between the hydrogen and the carbon atom to which the hydrogen binds. The parameters $\epsilon_{\alpha},\gamma_{\alpha,n}$ are taken from Table I of Ref.\onlinecite{ihnatsenka2011dirac}. As is discussed in Sections VI and VII of Ref.\onlinecite{ihnatsenka2011dirac}, the values of $\epsilon_{\alpha},\gamma_{\alpha,n}$ were chosen so as to accurately describe the scattering of graphene $\pi$ electrons by the $sp_3$ bonded hydrogen/carbon complex, including the effects of the carbon $p_x$, $p_y$ and $s$ valence orbitals, and agree with the results of DFT calculations \cite{Wehling10}.  As well as hydrogen on pristine graphene (the case $\Delta_n=0$), we also consider hydrogen on graphene with broken inversion symmetry in each unit cell,
(as in graphene on h-BN) with $\Delta_n=\pm 0.0602$ eV.\cite{George11} In that case, $\Delta_r$ and  $\Delta_l$, the site energies for carbon atoms under right and left hand lines of H atoms, have the opposite signs, breaking the left-right mirror symmetry of the nanostructure.

\section{Conductance Calculations} 

To explore the effects of the hydrogen lines  on electron transport, we have carried out 2-terminal conductance calculations using a fully quantum mechanical approach within Landauer theory. According to the Landauer formula of electron transport, the electrical conductance $G$ at zero temperature in the linear response regime is\cite{Econ81,Fish81,ReviewRibbon}
\begin{equation}\label{conductance}
G=G_{\circ}\sum_{i,j}T_{i,j}(E_\text{F}),
\end{equation} 
where $G_{\circ}=\frac{2e^2}{h}$ and $T_{i,j}(E_\text{F})$ is the transmission probability of electron from lead $j$ in the source contact to lead $i$ in the drain contact at the Fermi energy $E_\text{F}$. We calculate $T_{i,j}(E_\text{F})$ by solving the Lippmann-Schwinger equation for electron scattering through the nanostructure as is explained in detail in Appendix A of Ref. \onlinecite{azari2017gate}. 
 \begin{figure*}[t]
\centering
\includegraphics[width=1.0\linewidth]{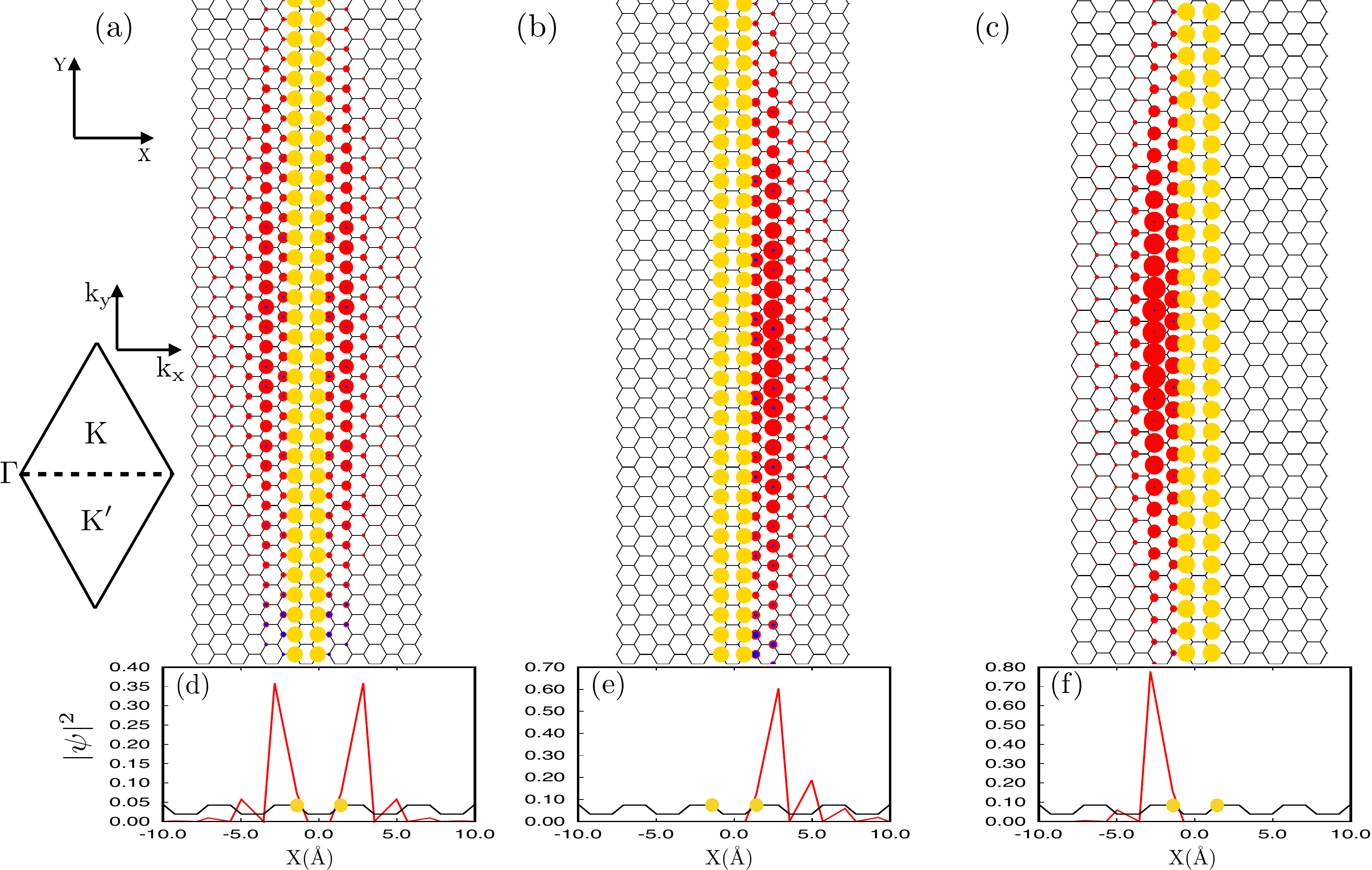}
\caption{(Color online) (a-c) Calculated electron accumulation (i.e., current-induced electron population)   $A_n^{K'(K)}$ in valley $K'$ ($K$) near H lines in Fig.\ref{nanostructure}  shown as red (blue) disks. (a): Symmetric case for $E_\text{F} = -0.0065$eV.   (b) and (c): Asymmetric case for $E_\text{F} = 0.0502$ and $-0.0638$eV (red and blue H-bands in Fig.\ref{conband} (d)), respectively. Electrons flow in the positive $y$-direction. Disk diameters proportional to $A_n^{K'(K)}$, but diameters in (c) are scaled down by factor 7 relative to those in (a) and (b). (d), (e), (f): Calculated square amplitudes (red) of representative electron H-band eigenstates of graphene nanoribbons near lines of hydrogen corresponding to the valley accumulations of (a), (b), (c), respectively. Ribbon unit cell fragment is also shown. Inset: Rhombic Brillouin zone of graphene with Dirac points $K$ and $K'$.}
\label{mapstate}    
\end{figure*}     

The calculated 2-terminal conductances [Eq.\ref{conductance}] of the quantum dot in Fig.\ref{nanostructure}(a) for the symmetric ($\Delta_n=0$) and asymmetric ($\Delta_n \ne 0$) cases are shown in Fig.\ref{conband} (a) and (b), respectively. The results for the graphene quantum dots with and without the hydrogen lines present are shown in blue and red. The calculated conductances for the symmetric and asymmetric cases are almost identical if the Fermi energy $E_\text{F}$ is far from zero. However, if the hydrogen lines are present, they differ markedly for Fermi energies around zero. In the symmetric case (blue curve, Fig.\ref{conband} (a)), a striking enhancement of the conductance to $G\sim 4e^2/h$ occurs when the Fermi energy lies in the range $-0.0157\text{eV}<E_\text{F}<-0.005\text{eV}$.  For the same model parameters, the Dirac point resonance for a single hydrogen atom on graphene is centered at $-0.00702$eV\cite{ihnatsenka2011dirac}, which is in this range. This suggests that the conductance enhancement for Fermi energies around zero is due to transmission of electrons from the source to the drain mediated by Dirac point resonances of the individual hydrogen atoms which together induce a conducting channel across the quantum dot. We then attribute the series of conductance maxima and minima in the range $-0.0157\text{eV}<E_\text{F}<-0.005\text{eV}$ seen in the inset of Fig.\ref{conband} (a) to multiple reflections of electrons in this channel at the ends of the lines of hydrogen atoms. The much smaller conductance near zero energy in the absence of hydrogen (shown in red in Fig.\ref{conband}(a)) is due to quantum tunneling in the energy gap around zero energy that results from quantum confinement due to the finite size of the pristine graphene quantum dot. If the mirror symmetry is broken with respect to the hydrogen lines by modifying the on-site energies of carbon atoms as discussed above, two peaks (with differing fine structures) occur near  $\sim 0.05$ and -0.06 eV, as shown by the blue curve and insets in Fig.\ref{conband}(b). 
\section{The Role of Band Structure}
To clarify the underlying physics, in Fig.\ref{conband} (c) and (d) we have plotted the band structures of infinite uniform graphene nanoribbons with and without double lines of adsorbed hydrogen atoms. The lines of hydrogen atoms on the ribbons are infinite, run along the centers of the ribbons, and are oriented in the $y$-direction, as in Fig.\ref{nanostructure}. This implies that the ribbons have zigzag edges. As is seen in Fig.\ref{conband} (c), the presence of the hydrogen lines in the mirror symmetric case results in a four-fold degenerate (including spin) narrow band close to zero energy (red) in addition to the flat band at zero energy\cite{nakada1996edge,fujita1996peculiar} (taupe colored) that is due to the zigzag ribbon edges. Fig.\ref{conband} (d) shows that breaking the mirror symmetry (with $\Delta_r=-\Delta_l$) opens a band gap of width $\sim 2|\Delta|$ around zero energy, splitting both the flat band due to the zigzag edges (taupe) and the band due to the hydrogen (red and blue). It is evident that states of these hydrogen-induced bands give rise to the conductance peaks near the Dirac point in Fig.\ref{conband} (a) and (b). Note that there are no features due to zigzag edges in the conductance plots in Fig.\ref{conband}(a) and (b) because the quantum dots considered there have only armchair edges as in Fig. \ref{nanostructure}.

In Fig.\ref{conband} (c) and (d) the hydrogen-induced bands have positive (negative) velocities $v=\frac{1}{\hbar}\frac{dE}{dk}$ for negative (positive) values of $k$ that can be regarded as projections of valley $K'$ ($K$) on the $y$-axis. [$K'$  and $K$ are defined in the insets of Fig.\ref{mapstate}. The corresponding ribbon $k$ values are indicated at the top of Fig.\ref{conband} (c),(d)]. This suggests that H-band electrons traveling in the positive (negative) $y$-direction in Fig.1 belong to valley $K'$ ($K$). Consequently, the Dirac point resonance states induced by the lines of H-atoms should function as an effective valley filter. Also an electric current mediated by the Dirac point resonances [i.e., for $E_\text{F}$ at a conductance peak near the Dirac point in Fig.\ref{conband}(a) and (b)] should induce a strong valley polarization near the adsorbed hydrogen lines. These heuristic ideas are fully confirmed by our calculations of the valley accumulations induced by electric currents that we present next.

\section{Valley Filtering and Valley Accumulation}  

To this end, we have solved the Lippmann-Schwinger equation to find the electronic scattering eigenstates $|\psi^l\rangle$ of the coupled system (graphene-hydrogen and leads) in Fig. \ref{nanostructure}. These scattering eigenstates were then projected onto the Bloch states belonging to the two graphene valleys. (Our method of solving Lippmann-Schwinger equations and carrying out valley projections is explained in Appendices A and B of Ref.\onlinecite{azari2017gate}). Then for a nanostructure with two contacts each at a specific electrochemical potential $\mu_i$, $A_n^{K(K')}$, the current-induced electron accumulation in valley $K$ $(K')$ calculated at atomic site $n$ is given by\cite{azari2017gate}:
\begin{equation}\label{accumulation}
A_n^{K(K')}=\frac{1}{2\pi}\sum_{l,i}|\langle n|\psi^l_{K(K')}\rangle|^2\frac{\partial \zeta^l}{\partial E}\Delta\mu_i    
\end{equation}
where $|\psi_K^l\rangle$,$|\psi_{K'}^l\rangle$ are the projections of the scattering state $|\psi^l\rangle$ on the valleys $K$ and $K'$, $|n\rangle$ is the $p_z$ atomic orbital at site $n$, and ${\Delta\mu}_i$ is the electrochemical potential difference between contact $i$ and the contact with the lowest electrochemical potential. 
Here the scattering state $|\psi^l\rangle$ emanates from semi-infinite 1D ideal lead $l$ represented by a tight-binding chain such that on site $m$ of the chain $\langle m|\psi^l\rangle = e^{i \zeta^l m} + r^l e^{-i \zeta^l m}$ where $r^l$ is the reflection amplitude of the incoming state $|\psi^l\rangle$ from the nanostructure back into ideal lead $l$, and $E$ is the energy eigenvalue corresponding to state $|\psi^l\rangle$.

Representative examples of the calculated spatial distributions of the current-induced electron accumulations $A_n^{K}$ and $A_n^{K'}$ in valleys  $K$ and $K'$ near the double hydrogen line in Fig. \ref{nanostructure} are shown as blue and red disks, respectively, in Fig.\ref{mapstate} (a-c). The disk diameters are proportional to $A_n^{K(K')}$. In each case shown, the Fermi level is in a hydrogen-induced band of states, i.e., a H-band of Fig.\ref{conband} (c) or (d).  The results for the symmetric ($\Delta=0$) case are in Fig.\ref{mapstate} (a). Those for the asymmetric ($\Delta_r=-\Delta_l$) case are in Fig.\ref{mapstate} (b) and (c) for $E_\text{F}$ in the upper and lower H-band respectively. In each case the electron flow is in the positive $y$-direction. In all cases the accumulation is overwhelmingly in the $K'$ valley (red disks dominate), as is predicted by our heuristic reasoning presented above. I.e., electron flow in the positive 
$y$-direction is due to $K'$ valley electrons for which the velocity in the H-bands in Fig.\ref{conband} is positive. We attribute the much smaller population of $K$ valley electrons (blue disks) to the minority of electrons traveling in the negative $y$-direction after partial reflection at the upper ends of the hydrogen lines in Fig.\ref{nanostructure}. If we define the valley filter efficiency as the current-induced valley polarization $=\frac{\sum_n A_n^{K'}}{\sum_n (A_n^K+A_n^{K'})}$, then we find efficiencies of 90.8$\%$ in Fig.\ref{mapstate}(a) [symmetric case], and 91.8 and 95.1$\%$ in Fig.\ref{mapstate}(b) and (c) [asymmetric cases], respectively. It is possible to tune the filter efficiency by modifying conditions so as to vary the electron reflection probability.  For example, locating $E_\text{F}$ in a broader conductance peak (in an inset of Fig.\ref{conband} (a) or (b)) lowers the reflection probability and, accordingly, lowers the $K$ valley electron accumulation and raises the filter efficiency.  
The calculated wave function distributions over the unit cell of a graphene nanoribbon with a double hydrogen line, Fig.\ref{mapstate} (d-f), confirms that the hydrogen-induced states are exponentially localized around the hydrogen lines in a similar way to the corresponding valley accumulations in Fig.\ref{mapstate} (a-c).

In all of the above cases the valley filtering allows transport of valley $K'$ electrons while excluding almost all valley $K$ electrons. However, if the direction of the electric current is reversed, the filtering switches to favoring valley $K$ instead of valley $K'$.   
It should be noted that, this novel valley filtering mechanism only operates at the energies of the hydrogen-induced Dirac point resonance bands. It does not function if the Fermi energy is far from the Dirac point resonances.

\section{Conclusions} 

In conclusion, the present work has revealed an efficient, experimentally realizable valley filtering mechanism that exploits in a novel way the Dirac point resonances due to arrays of hydrogen atoms adsorbed on graphene  quantum dots. We have shown that a double row of adsorbed hydrogen atoms induces an electrically conducting channel in the graphene for Fermi energies near the Dirac point and within the quantum confinement-induced energy gap of the corresponding pristine graphene quantum dot. We predict that an electric current passing through the channel in a given direction is carried by electrons of {\em only one} of the two graphene valleys. If the direction of the current is reversed, conduction switches to electrons of the other valley. This novel valley filtering mechanism applies for hydrogen adsorbed on pristine graphene and also on graphene whose symmetry is broken by a h-BN substrate. We have shown that if electron reflection at the ends of the channel is minimized, electrons belonging almost entirely to just one graphene valley accumulate in a nanometer-wide region along the conducting channel. 
Thus we have proposed here a novel, efficient valley filter, accumulator and switch. 
\begin{acknowledgments}
This work was supported by NSERC, Westgrid, CIFAR, and Compute Canada.
\end{acknowledgments}                               

{

\end{document}